\documentstyle[psfig]{caosp}
\begin{document}
\htitle{Light element abundances in He-rich stars}
\hauthor{M. Zboril and P. North}
\title{Light element abundances in He-rich stars}
\author{M. Zboril \inst{1} \and P. North\inst{2}}
\institute{Armagh Observatory, Armagh, College Hill, BT61 9DG, N. Ireland \and
Institut d'Astronomie de l'Universit\'{e} de Lausanne,
CH-1290 Chavannes-des-Bois,Switzerland}
\date{December 22, 1997}
\maketitle
\begin{abstract}
We present an abundance analysis
of light elements in He-rich stars. The analysis is based on both low and high 
resolution observations collected at ESO, La Silla, Chile in the optical region
and includes 6 standards and 21 He-rich stars. Light-element abundances
display a diverse pattern: they range from under-solar up to above-solar values.
\end{abstract}

\section{Introduction}
The He-rich stars are the most massive CP stars. Their helium
abundance ranges from nearly solar up to larger
than unity with respect to hydrogen ($n(He)$=1.). Spectroscopic and
photometric variability is explained by an abundance distribution
across the stellar surface. Given standard atmospheric conditions in
B type stars (assuming no wind), diffusion is unable to support helium in the
stellar atmosphere and helium sinks. However, Vauclair (1975) showed that
diffusion could lead to He overabundance in the presence of mass-loss.
Models of abundance anomalies with selective mass loss for He-rich stars suggest normal CNO abundances as a test (Michaud et al. 1987). So in this 
contribution, we analyse both low and high resolution 
ESO spectra to obtain light element abundances for He-rich stars and
to put constraints on the theoretical model.
He and preliminary CNO abundances have already been examined
by Zboril et al. (1994, 1997).
 
\section{Observations}
High resolution spectra (resolving power $R=30000$) were obtained using
the CAT telescope and the CES spectrograph at ESO, La Silla, Chile,
in April 20-24, 1992 and in June 25-29, 1993.
The low resolution spectra (resolving power $R\sim 4150$) were obtained using 
the 1.5m spectrographic telescope and the Boller \& Chivens spectrograph at ESO
in January 4-11, 1993. The total sample consists of 6 standard stars and 21
He-rich stars.

\section{Abundance analysis and results}
Given the instrumentation set up the abundance analysis from high resolution 
spectra is based on the range 4235-4270\AA~while low resolution spectra 
cover the 3952-4938\AA~interval. Thus the optical region is fairly well covered;
however, photospheric abundances are in general difficult to determine due to
the weakness of the majority of photospheric lines. The UV resonance lines are
much more pronounced but require exhaustive NLTE treatment.
To start with, we derived carbon abundance and $v\sin i$ values from the
4268\AA~transition. The following analysis is based on the best atmosphere 
models for early type stars, i.e. Kurucz (1992) models. New atmosphere models 
for He-rich stars reflecting helium abundance, proper line blanketing and
NLTE treatment are being built up.

It turned out that it was better to put aside the constant for Stark broadening 
for the C\,{\sc ii} 4268 transition and to apply instead the classical
formula, $\Gamma_{S} \sim n_{e}.n_{eff}^{5}$, where $n_{e}$ stands for
electron plasma density and $n_{eff}$ for the effective quantum number of the
upper level; otherwise, the line wings did not fit any of the observed line 
profiles, especially for sharp-lined stars. Other damping constants (e.g. 
natural radiative) associated with broadening mechanisms matched the 
observations very well. Other line parameters, oscillator strengths etc., have 
been maintained from the original line list.
 The abundance analysis has been performed by a
standard synthetic spectrum method. The number of well pronounced
transitions for light elements in the spectrum is not large enough to estimate
reliably microturbulent velocity and 
therefore we adopted a depth-independent microturbulent velocity value
of 2 km\,s$^{-1}$. The abundances (defined relative to hydrogen)
for light elements can be found in Table 1. The column
entitled ``method'' stands for effective temperature determination method.
Note that non-solar magnesium abundance often coincides with helium 
overabundance; the magnesium line is in the wing of a helium spectral line and 
consequently is formed in higher atmospheric layers.
In the error analysis, the errors on the continuum 
location and total equivalent width determination have been considered as the
maximal error contribution for low resolution spectra, and amount to 28\% of 
the derived abundance. In the case of high-resolution spectra, error bars were 
computed using standard analysis, assuming error propagation from several 
sources. In this case the total abundance error does not exceed 11\%.
\section{Conclusion}
The observed CNO abundances do not entirely fulfill the predictions of the
model proposed by Michaud et al. (1987). In particular, C appears underabundant
in most He-rich stars (in agreement with Hunger \& Groote 1993), especially the 
hotter ones. More detailed models including the magnetic and wind geometry
would be welcome.

\begin{table}[t]
\small
\begin{center}
\caption{Elemental abundances $n(X)/n(H)$ in He-r stars (upper part of the 
table) and in normal, reference stars (lower part). The solar abundances are 
listed at the very bottom. The DM numbers obey the HD rule.}
\begin{tabular}{|c|r|r|r|r|r|r|r|}\hline
HD/DM & T$_{eff}$ & method & C & N & O & Mg & $v\sin i$ \\ \hline
169467 & 16600 & phot. & 4.2e-4 & 2.2e-4 & 6.7e-4 & --- &30. $\pm$3.\\
168785 & 23400 & phot. & 2.4e-5 & 4.0e-5 & 2.7e-4 & --- &14. $\pm$2.\\
-69 2698& 25300& phot. & 7.0e-5 & 1.1e-4 & 9.7e-5 & ---& 30. $\pm$3 \\
149257 & 24900 & phot. & 7.2e-5 & 7.7e-5 & 1.5e-4 & --- &40. $\pm$4\\
133518 & 18600 & spectr.& 2.3e-4& 2.2e-4 & 3.3e-3 & 8.0e-6&0. $\pm$1.\\
96446  & 22000 & spectr.& 4.5e-5& 1.1e-4 & 1.1e-3 & 8.0e-6&0. $\pm$1.\\
92938  & 15000 & spectr.& 9.2e-4& 1.1e-4 & ---    & 3.4e-5&125. $\pm$4.\\
-46 4639&22000 & spectr.& 1.2e-4& 1.1e-4 & 4.4e-3 & 3.4e-5&80. $\pm$4.\\
66522  & 18800 & spectr.& 3.7e-4& 3.3e-4 & 5.5e-3 & 8.2e-6&0. $\pm$1.\\
108483 & 19100 & spectr.& 2.2e-4& 1.1e-4 & 6.7e-4 & 3.4e-5& 130$\pm$12.\\
-62 2124&26000 & spectr.& 1.1e-4& 1.1e-4 & 6.7e-4 & 3.4e-5& 75$\pm$8.\\
60344  & 21700 & spectr.& 7.0e-5& 3.3e-4 & 2.1e-3 & 3.4e-5&55$\pm$6.\\
64740  & 22700 & spectr.& 8.0e-5& 3.3e-4 & 2.8e-3 & 3.4e-5&130$\pm$9.\\
58260  & 19000 & spectr.& 9.0e-5& 2.2e-4 & 1.4e-3 & 3.4e-5&45$\pm$6.\\
-27 3748&22700 & spectr.& 3.0e-5& 1.1e-4 & 6.7e-4 & 7.0e-6&45$\pm$4.\\
264111 & 23200 & spectr.& 2.0e-5& 1.9e-4 & 1.0e-3 & 3.4e-5&75$\pm$8.\\
260858 & 19200 & spectr.& 8.0e-5& 2.2e-4 & 1.4e-3 & 1.5e-5&47$\pm$3.\\
37776  & 21800 & spectr.& 8.0e-5& 2.2e-4 & 1.4e-3 & 3.4e-5&75$\pm$7.\\
37479  & 22200 & spectr.& 1.7e-5& 1.1e-4 & 6.7e-4 & 
3.4e-5&$\sim$100$\pm$15.\\
37017  & 19200 & spectr.& 2.7e-5& 9.0e-5 & 6.0e-4 & 1.8e-5&$\sim$80.$\pm$9.\\
36485  & 18400 & spectr.& 5.0e-4& 1.1e-4 & 6.7e-4 & 3.4e-5&54$\pm$3.\\ \hline
144218 & 20300 & phot.  & 2.4e-4& 3.3e-4 & 1.3e-3 & ---& 60. $\pm$4.\\
122980 & 19400 & phot.  & 4.4e-4& 4.4e-4 & 4.3e-3 & 3.4e-5& 20. $\pm$2.\\
110879 & 19700 & phot.  & 6.2e-4& 2.2e-4 & ---    & 3.4e-5& 130. $\pm$4.\\
56139  & 18000 & phot.& 1.2e-4& $\sim$1.1e-4&$\sim$6.7e-4& 3.4e-5& 
80.$\pm$7.\\
105435 & 26000 & phot.& 2.2e-4&  $\sim$1.1e-4&$\sim$6.7e-4& 
---&$\sim$240.$\pm$11.\\
121790 & 19600 & spectr.& 2.2e-4& 2.2e-4 & 1.4e-3& 3.4e-5&90.$\pm$8.\\ \hline
solar  & ---   & ---    & 3.7e-4& 1.1e-4 & 6.7e-4 & 3.4e-5&---\\ \hline
\end{tabular}
\end{center}
\end{table}

\end{document}